\documentstyle[preprint,aps]{revtex}
\begin{document}
\preprint{\begin{tabular}{c}
\hbox to\textwidth{hep-ph/9702355 \hfill BROWN-HET-1065}\\[-10pt]
\hbox to\textwidth{               \hfill YUMS-96-30}\\[-10pt]
\hbox to\textwidth{                 \hfill SNUTP 96-116}\\[-10pt]
\end{tabular}}
\draft
\title{ Higgs Boson Mass Bounds in the Standard and  Minimal 
       Supersymmetric Standard Model with Four Generations}
\author{ Sin Kyu Kang}
\address {\it Department of Physics, Brown University, 
 Providence, RI 02912, USA} 
\author{ Gye~T.~Park}
\address {\it Department of Physics, Yonsei University,
 Seoul, 120-749, Korea}
\maketitle
\begin{abstract}
We study the question of distinguishability of the Higgs sector between
the standard model with four generations(SM4) and the minimal supersymmetric standard model with four
generations (MSSM4).
We find that a gap exists between the SM4 and MSSM4 Higgs boson masses for
a range of the fourth generation fermion mass considered in the analysis at 
a fixed top quark mass.
We also compare the Higgs boson mass bounds in these models with those in the 
standard and  the minimal supersymmetric standard models.
\end{abstract}

\newpage
With the discovery of the top quark at the Tevatron \cite{top}, 
the Higgs boson is now the only unknown sector in the context of the standard 
model (SM).
Despite the remarkable successes of the SM in its excellent agreement with
the precision measurements at present energies\cite{ewdata}, it is generally believed
that the SM is not the final theory of elementary particle interactions.
The minimal supersymmetric standard model (MSSM) \cite{susy} is one of the most popular
extensions of the SM.
Because of the nature of supersymmetry (SUSY), the Higgs sector in the MSSM
consists of two CP-conserving Higgs doublets with opposite hypercharge.

In the SM, the Higgs boson mass is usually considered as an adjustable parameter
because the quartic coupling of the Higgs potential is arbitrary.
Nevertheless if certain theoretical assumptions are imposed,
upper and lower bounds on the Higgs boson mass can be obtained.
The requirement of the vacuum stability yields a severe lower bound on the Higgs boson mass which 
depends on the top quark mass and the cut-off scale beyond which the SM is
no longer valid \cite{cab,vac}, while an upper bound follows from the requirement that no
Landau singularity appears up to a scale\cite{triv}.
In the MSSM,
an intrinsic upper bound on the lightest Higgs boson mass is obtained
from the quartic Higgs coupling which is no longer arbitrary but is constrained 
by SUSY \cite{susyup,susyup2}.

On the other hand, since there is still no experimental evidence for the 
absence of extra generations, it would also be interesting to study  how
the Higgs boson mass is limited in the presence of extra generations.
In the SM with extra generations, several authors \cite{fourge,novi,nn} have derived the upper and
lower bounds on the Higgs boson mass as functions of the extra fermion masses.
In the MSSM with four generations (MSSM4), the bounds on the lightest Higgs 
boson mass have been calculated very recently in Ref.\cite{skk}.

In view of the search for the Higgs boson at future accelerators such as LEP200 and LHC,
it would be worthwhile for one to examine the bounds on the Higgs boson in the SM and
its extensions and to look for any distinctive features since an actual measurement of the Higgs boson
mass could serve to exclude or at least constrain  some of the models for the electroweak
symmetry breaking.
This is the aim of this letter.
In this work, we will pay a special attention to the question of distinguishability
between the SM with four generations (SM4) and MSSM4 Higgs bosons as well
as between the SM/MSSM and SM4/MSSM4 Higgs sectors.
In order to do this we adopt a basic assumption \cite{assume}that all super partners of the
SM particles and another Higgs scalar orthogonal to the lightest one have masses
of order of the supersymmetry breaking scale $M_{susy}\geq 1$ TeV.
Then, the effective low-energy theory below $M_{susy}$ is equivalent to the SM
with a single Higgs doublet and three or four generations.

Since the bounds on the Higgs boson mass in the models with four generations 
would depend on the fourth generation fermion masses, 
it is convenient to impose the possible constraints on 
the fourth generation fermion masses.
The recent precision tests of the SM \cite{ewtest}
and the good agreement between the direct measurements of the top quark mass
at the Tevatron \cite{top} and its indirect determination from the global fits to 
the electroweak data \cite{ewtest,my,gents}
demonstrate that  no significant violation of the isospin symmetry 
for the extra generations is observed.
Thus the masses of the fourth generation isopartners must be highly  degenerate
(see e.g. \cite{nn}).
To reduce the number of parameters we will consider the fourth generation
fermions with the common mass $m_4$.
Recently, the limit on the masses of the extra neutral and charged leptons,
$m_N$ and $m_E$, has been improved by LEP1.5 to
$m_N > 59$ GeV and $m_E> 62 $ GeV \cite{lep15}.
On the other hand, the upper limit on the fourth generation fermion masses
coming from the vacuum stability and the perturvative calculation is about
120 GeV for the scales $\Lambda \geq 10^{12}$ GeV at which new physics emerges 
\cite{conv,nn}.
Taking into account these observations, we will restrict
the range of $m_4$ to $50 \lesssim m_4 \lesssim 120 $ GeV in our analysis.
In addition, we will assume that the fourth generation quarks do not mix with the known quarks.
This is possible since the mixing angles are so small that the new particles can leave
the Tevatron detectors without decaying.

Now, let us focus on the calculation of the Higgs boson mass.
In order to evaluate the quantum effects on the Higgs boson mass
we adopt the renormalization group (RG) analysis.
The renormalization group equations (RGEs) for the gauge and Yukawa
coupling constants can be found in the literature \cite{skk,conv,rge}
Note that our calculation is up to one loop order.
In order to solve the RGEs numerically, we take the initial conditions
for the gauge coupling constants to be,
\begin{eqnarray}
 g_1(M_Z) = 0.3578  \\
 g_2(M_Z) = 0.6502  \\
 \alpha_s(M_Z) = \frac{g_3^2(M_Z)}{4\pi} = 0.12 .
\end{eqnarray}
Note that the uncertainties in our numerical results due to the experimental 
errors in $g_1(M_Z)$ and $g_2(M_Z)$ are negligible while it is not
for $\alpha_s(M_Z)$.
Therefore, we will discuss briefly on the uncertainty due to the experimental error in 
$\alpha_s(M_Z)$.
For the Yukawa couplings of the top quark and the fourth generation fermions, 
we impose the following mass relation as a boundary condition,
\begin{equation}
m_i=h_i(m_i)v/\sqrt{2},
\end{equation}
where $v=(\sqrt{2}G_F)^{-1/2}=246 $ GeV, $i=t, T, B, N$ and $E$, and the fourth
generation quarks and leptons are denoted by $(T,B)$ and $(N,E)$ respectively.
In order to calculate the Higgs boson mass $m_H$, we will use the following 
relation \cite{assume}
\begin{equation}
 \frac{m_H^2}{M_Z^2} = \frac{4\lambda(m_H)}{g_1^2(M_Z)+g_2^2(M_Z)}.
\end{equation}
Note that the above mass relations to be imposed as boundary conditions are
different from those in Ref.\cite{novi,nn}.
We will comment below on the effects of using different boundary conditions
on the Higgs boson mass.
In particular, when we calculate the lightest supersymmetric Higgs boson mass,
 we impose the following SUSY relation on the Higgs coupling
$\lambda (\mu)$ \cite{bc}
\begin{equation}
 \lambda (M_{susy}) = \frac{1}{4}(g_1^2(M_{susy})+g_2^2(M_{susy}))\cos^2 2\beta,
\end{equation}
where $\tan{\beta}$ is the ratio of the vacuum expectation values 
of the two Higgs scalars.
We also restrict the range of the top quark mass
to $160 \lesssim m_t \lesssim 180 $ GeV,
which covers within the current experimental errors the most part of the measured top quark masses from Tevatron.
In this work, we obtain the vacuum stability lower bounds on the SM(4) 
Higgs boson mass by
assuming that the SM(4) is valid all the way up to $10^{19}$ GeV, and for large
values of cutoff, $\Lambda \geq 10^{10} $ GeV, these bounds
depend only weakly on the values of $\Lambda $.

In Fig.1, we present the lower bounds on the SM4 Higgs boson mass and the upper
bounds on the lightest MSSM4 Higgs boson mass as a function of $m_4$ 
for $m_t= 160, 170, 180 $ GeV with (a) $M_{susy}=1$ TeV and
(b) $M_{susy}=10$ TeV.
We find that a gap exists between the SM4 and MSSM4 bounds.
The difference between  the SM4 and MSSM4 bounds also grows with $m_4$.
However, for smaller values of $m_4$ and higher SUSY-breaking scales,
the difference between the  bounds diminishes.
Since both SM4 and MSSM4 are extensions of the SM, it would be interesting
to compare the bounds on the SM4 and MSSM4 Higgs boson mass with those of SM and
/or MSSM.

In Fig.2, we show the bounds on the Higgs boson mass of SM(dotted), MSSM(dot-dashed),
SM4 (dashed) and MSSM4(solid) for $m_t=180 $ GeV and 
$M_{susy}=1 $ TeV.
As can be seen in the figure, the differences between the SM and SM4 Higgs boson
mass bounds are more pronounced than those between the MSSM and MSSM4 bounds.
In fact, this agrees with the results from Ref.\cite{skk} that only for large values of $m_4$ 
MSSM4 Higgs boson mass can be distinguished from the corresponding MSSM
Higgs boson mass for the fixed values of $m_t, M_{susy}$, and $\tan \beta $.
As one can see from Fig.2, there are four splitted regions of the Higgs boson
mass $m_H$ for $m_t=180$ GeV and $M_{susy}=1$ TeV : \\
\begin{eqnarray*}
  (i)~~~~  160 - 195 ~~\mbox{GeV} \lesssim & m_H &\\
  (ii)~~~~~~~~~~~~ 155 ~~\mbox{GeV} \lesssim & m_H &\lesssim 
               160 - 195 ~~\mbox{GeV} \\
  (iii)~~~~~~~~~~~~125 ~~\mbox{GeV} \lesssim & m_H &\lesssim 155 ~~\mbox{GeV} \\
  (iv)~~~~~~~~~~~~~~~~~~~~~~~~~~~~  & m_H &\lesssim 125 ~~\mbox{GeV}
\end{eqnarray*}
The discovery of the Higgs in the regions $(i)$ and $(ii)$ would exclude
the MSSM with three or four generations for $m_t=180$ GeV.
Although the mass differences between the Higgs in both models increase as $m_4$ increases,
it is yet somewhat difficult for one to distinguish
between the SM and the SM4 Higg sectors with only a measurement of $m_H$
unless we have a strong constraint on $m_4$.
If the Higgs boson mass were to be in the region $(iii)$, another 
extension of the SM 
might be responsible for the mechanism of electroweak symmetry breaking.
For the mass region $(iv)$, the Higgs detection would serve as an evidence
for the supersymmetry.

Now, we ought to discuss here the theoretical uncertainties in our calculations.
Our numerical results have been obtained by integrating only the one-loop RGEs.
For consistency, we have also ignored the threshold corrections at the SUSY scale whose
effects are of the same order of magnitude as of the neglected two-loop contributions
to the RGEs when we calculate the lightest supersymmetric Higgs boson masses.
If we include two-loop contributions to the RGEs and maximal SUSY scale threshold corrections,
the lightest supersymmetric Higgs boson masses can increase by about $10-15 \%$.
Due to the different RG boundary conditions from those of Ref. \cite{novi,nn},
our numerical values for the vacuum stability lower bound on the SM4 Higgs 
boson mass are higher by  about 5-8 GeV than those of Ref. \cite{novi,nn}
for the ranges of $m_4$ amd $m_t$ 
considered.
There is also an uncertainty due to the experimental error in the strong coupling constant.
The shift of $m_H$ due to $\Delta \alpha_s(M_Z)= \pm 0.005$ is up to about 5 GeV.
Accordingly, more genuine improvements in the Higgs boson mass bounds are 
needed in view of the search for the Higgs boson at future accelerators.

In conclusion, we have examined the question of distinguishability of 
the Higgs sector between the SM4 and MSSM4.
We find that a gap exists between the SM4 
and MSSM4 Higgs boson masses for the fourth generation mass of the range
of $50 \lesssim m_4 \lesssim 120 $ GeV.
We have also compared the lower bound on the SM4 Higgs boson mass and the upper
bound on the lightest MSSM Higgs boson mass with the corresponding bounds in
the SM and the MSSM.

\acknowledgements

This work has been supported in part by
NON DIRECTED RESEARCH FUND, Korea Research Foundation (G.T.P), in part by the Basic Science Research Institute Program, the Ministry of Education Project No. BSRI-96-2425 (G.T.P), and in part by 
the Korea Science and Engineering Foundation (KOSEF)
through the SRC program of SNU-CTP (G.T.P).

In addition, one of us (S.K.K.) would like to acknowledge KOSEF for
a Post-doctoral Fellowship and also thank the High-Energy Theory Group,
Brown University for kind hospitality and support.

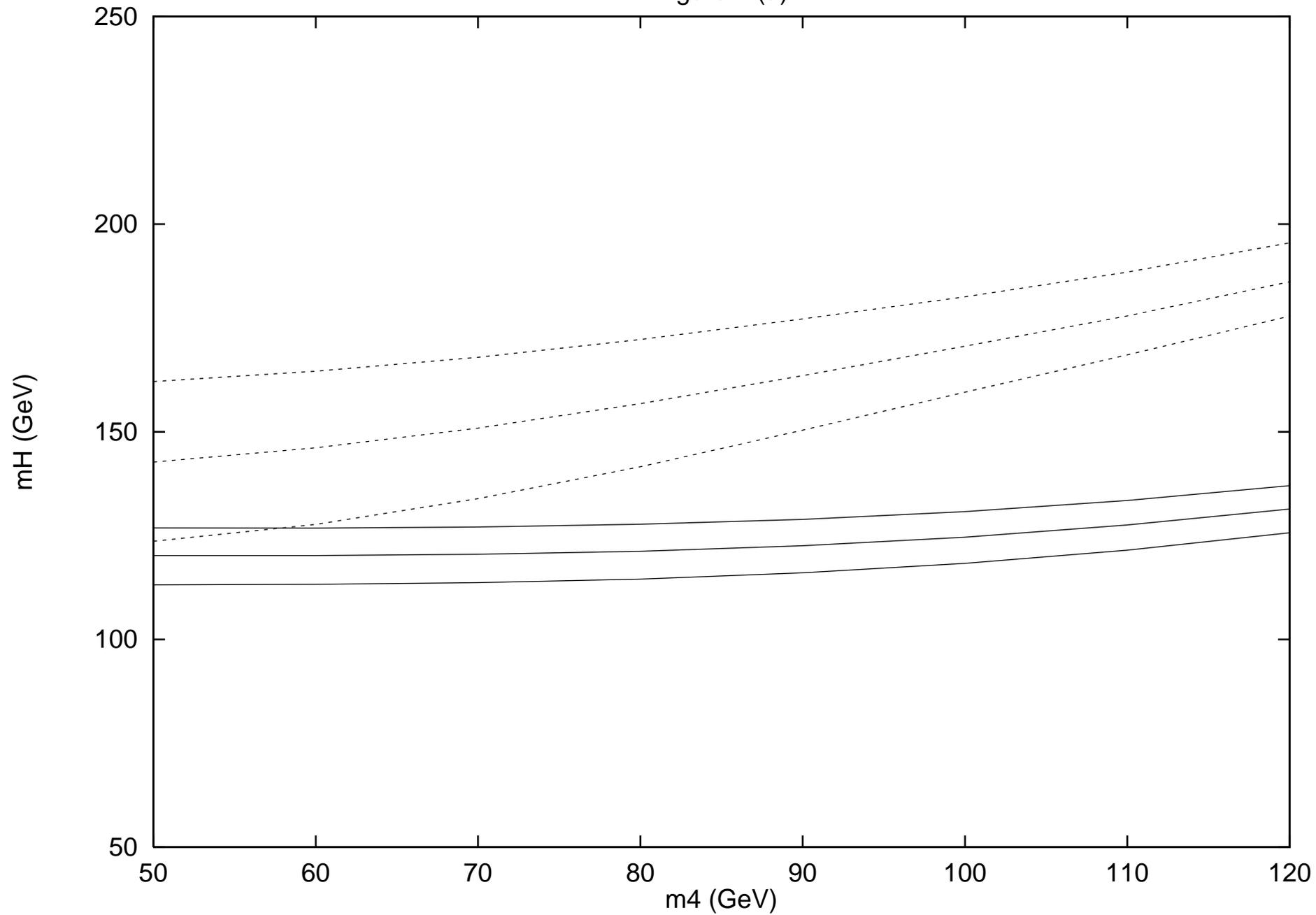
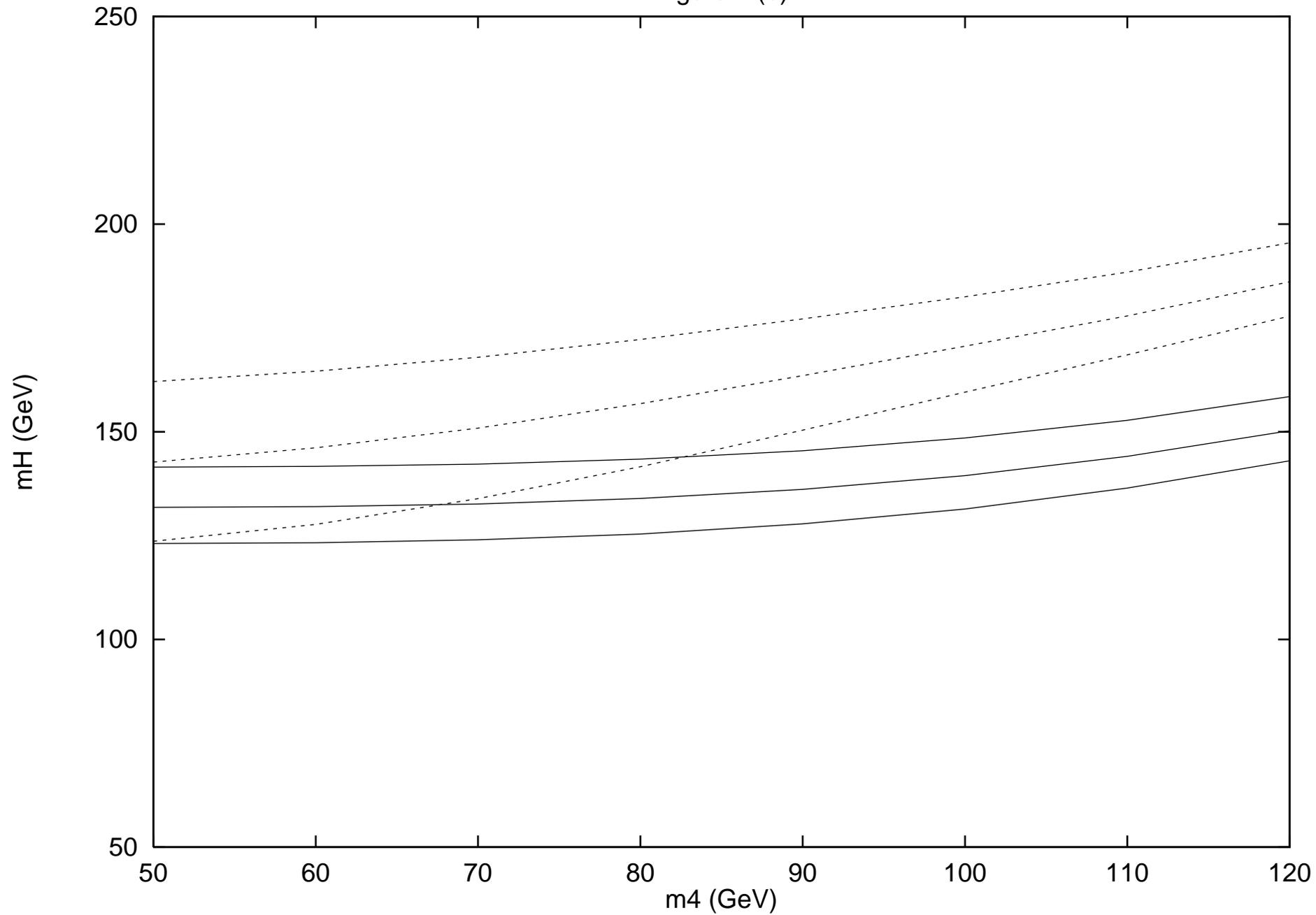
\begin{figure}
\caption
{ The lower bounds on the SM4 Higgs boson mass (dotted lines) and the upper
 bounds on the lightest MSSM4 Higgs boson mass (solid lines) as a function of $m_4$ 
for $m_t = 160, 170, 180$ GeV from below with (a) $M_{susy}=1$ TeV and 
(b) $M_{susy}=10$ TeV.}
\end{figure}
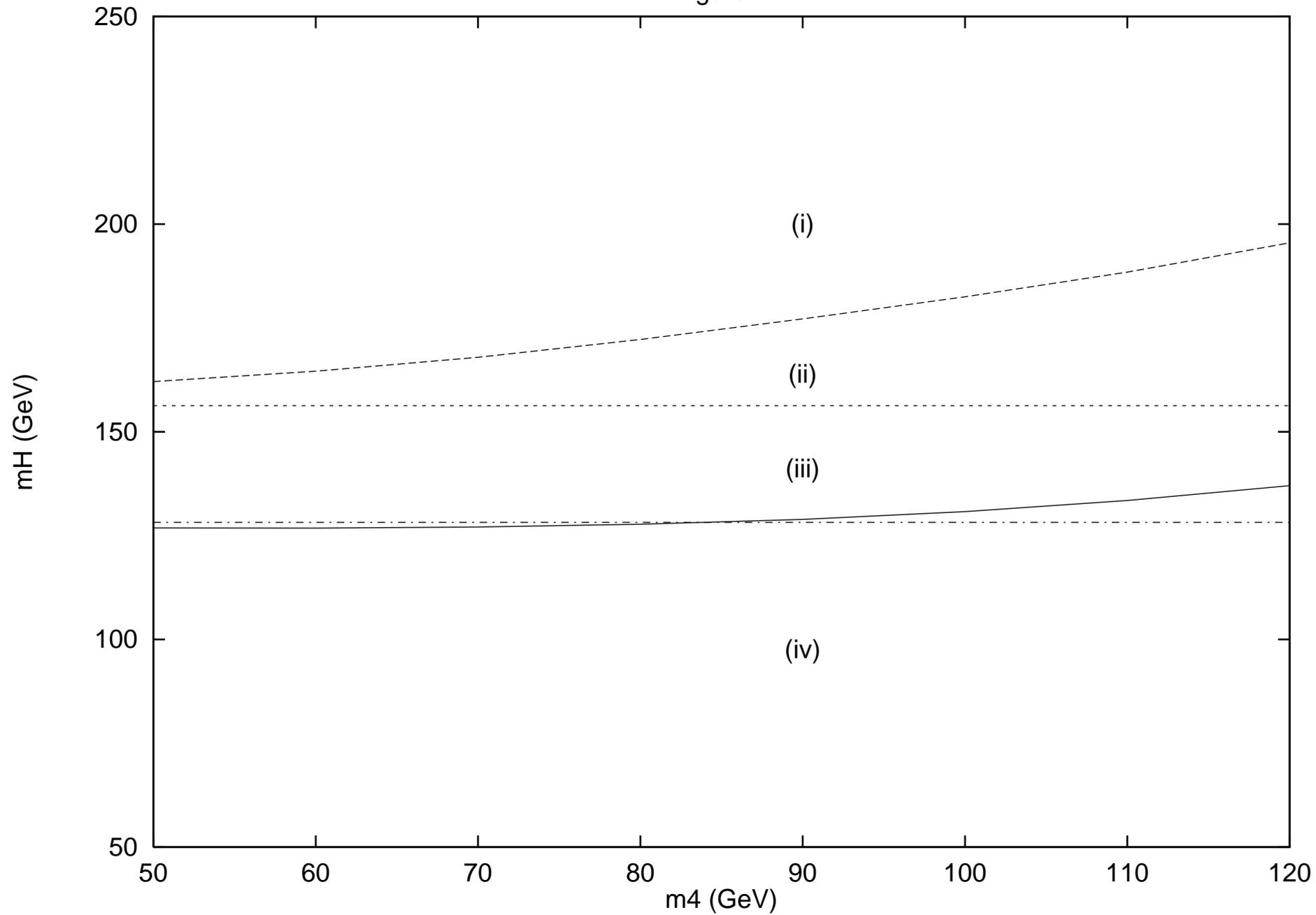
\begin{figure}
\caption
{ The Higgs boson mass bounds in SM (dotted), SM4 (dashed), MSSM
 (dot-dashed) and MSSM4 (solid) as a function of $m_4$ for
  $m_t = 180$ GeV  and $M_{susy}=1$ TeV.}
\label{2}
\end{figure}

\begin{references}
\bibitem{top} CDF Collaboration, F. Abe et al., Phys. Rev.Lett. {\bf 74},
             2676 (1995); D0 Collaboration, S. Abachi et al., Phys. Rev.
             Lett. {\bf 74}, 2632 (1995).
\bibitem{ewdata} The LEP Electroweak Working Group and the SLD Heavy Flavour
             Group, Data presented at the 1996 Summer Conferences,
             LEPEWWG/96-02.
\bibitem{susy} For reviews see  H. P. Nilles, Phys. Rep. {\bf 110}, 1 (1984);
             H. E. Haber and G. Kane, Phys. Rep. {\bf 117}, 76 (1985).
\bibitem{cab} N. Cabibbo, L. Maiani, G. Parisi, and R. Petronzio,
              Nucl. Phys. {\bf B 158}, 295 (1979).
\bibitem{vac} M. Sher, Phys. Rep. {\bf 179}, 273 (1989); Phys. Lett. 
             {\bf B 317}, 159 (1993); Phys. Lett. {\bf B 331}, 448 (1994);
              G. Altarelli and G. Isidori, Phys. Lett. {\bf B 337}, 114 (1994);
              J. A. Casas, J.R. Espinosa, and M. Quir\'{o}s,
	       Phys. Lett. {\bf B 342}, 171 (1995); J.R. Espinosa and 
               M. Quir\'{o}s, Phys. Lett. {\bf B 353}, 257 (1995). 
\bibitem{triv} M. Lindner, Z. Phys. C{\bf 31}, 295 (1986); 
              M. Sher, Phys. Rep. {\bf 179},273 (1989) and see also
              N. Cabibbo et al. (Ref. \cite{cab}).
\bibitem{susyup} H. E. Haber and R. Hempfling, Phys. Rev. Lett. {\bf 66}, 
               1815 (1991); Phys. Rev. {\bf D 48}, 4280 (1993);
               Y. Okada, M. Yamaguchi, and T. Yanagida, Prog. Theor. Phys. 
              {\bf 85}, 1(1991); Phys. Lett. {\bf B 262}, 54 (1991);
              J. Ellis, G. Ridolfi, and F. Zwirner, Phys. Lett. {\bf B 257}, 
              83 (1991); Phys. Lett. {\bf B 262}, 477 (1991);
              R. Barbieri, M. Frigeni, and F. Caravaglios, Phys. Lett. 
              {\bf B 258}, 167 (1991)
\bibitem{susyup2} R. Hempfling and A. H. Hoang, Phys. Lett. {\bf B 331}, 
                 99 (1994);
               J. Kodaira, Y. Yasui, and K. Sasaki, Phys. Rev. {\bf D 50},
                7035 (1994); 
              J. A. Casas, J. R. Espinosa, M. Quiros, and A. Riotto,
              Nucl. Phys. B{\bf 436}, 3 (1995); (E) {\bf B 439}, 466 (1995);
              M. Carena, J. R. Espinosa, M. Quiros, and C. E. M. Wagner, Phys. 
              Lett. {\bf B 355}, 209 (1995); M. Carena, M. Quiros, and 
              C. E. M. Wagner, Nucl. Phys.  {\bf B 461}, 407 (1996).
\bibitem{fourge} K. S. Babu and E. Ma, Z. Phys. {\bf C 29}, 45 (1985).
\bibitem{novi} H. B. Nielsen, A. V. Novikov, V. A. Novikov, and M. S. Vysotsky,
              Phys. Lett. {\bf B 374}, 127 (1996).
\bibitem{nn}  V. Novikov, talk given at 31th Recontres de Moriond:
              Microwave Background Anisotropies, Les Arcs, France, 16-23 Mar 
              1996, hep-ph/9606318.
\bibitem{skk} S. K. Kang, to appear in Phys. Rev. {\bf D 54}; see also
              K. Tabata, I. Umemura, and K. Yamamoto, Phys. Lett. {\bf B 129},
              80 (1983).
\bibitem{assume} See also Y. Okada et al. ( Ref.\cite{susyup}).
\bibitem{ewtest} J. Erler and P. Langacker, Phys. Rev. {\bf D 52}, 441 (1995);
             P. Langacker, NSF-ITP-95/40 and hep-ph/9511207;
             K. Hagiwara, KEK-TH-461 and hep-ph/9512425; K. Kang and S. K. Kang,
             Z. Phys. {\bf C 70}, 239 (1996); Z. Hioki, TOKUSHIMA 95-05 and 
             hep-ph/9511224;
            J. Ellis, G. L. Fogli, and E. Lisi, Z. Phys. {\bf C 69}, 627 (1996);
            CERN-TH/96-216 and hep-ph/9608329;
             P. H. Chankowski and S. Pokorski, Phys. Lett. {\bf B 356}, 
             307 (1995);
             S. Dittmaier, D. Schildknecht, and G. Weigleim, Phys. Lett. {\bf
             386}, 247 (1996); BI-TP 96/44 and hep-ph/9609488;
             W. Hollik, hep-ph/96010457.
\bibitem{my} See also K. Kang and S. K. Kang, in: Proc. Workshop on Quantum 
             Infrared Physics (Paris, June 1994), hep-ph/9412368; 
             in: Proc. Beyond the Standard Model IV (Lake Tahoe, December 1994),
             hep-ph/9503478 and references therein.
\bibitem{gents} T. Inami, T. Kawakami, and C. S. Lim, Mod. Phys. Lett.
              {\bf A 10}, 1471 (1995);
               V. Novikov, L. Okun, A. Rozanov, M. Vysotsky, and V. Yurov,
               Mod. Phys. Lett. {\bf A 10}, 1915 (1995);
               A. Masiero, F. Feruglio, S. Rigolin, and R. Strocchi, Phys. Lett.
               {\bf B 355}, 329 (1995).
\bibitem{lep15} LEP1.5 Collaboration, J. Nachtman, talk given at 31th Recontres 
                de Moriond: Electroweak Interactions and Unified Theories, 
                Les Arcs, France, 16-23 Mar 1996, hep-ex/9606015.
\bibitem{conv} J. W. Halley, E. A. Paschos, and H. Usler, Phys. Lett.
              {\bf B 155}, 107 (1985) and references therein; K. S. Babu and 
              E. Ma, Z. Phys. {\bf C 29}, 41 (1985).
\bibitem{rge} T. P. Cheng,  E. Eichten, and L. F. Li, Phys. Rev. {\bf D 9}, 
              2259 (1975);
              L. Maiani, G. Parisi, and R. Petronzio, Nucl. Phys. {\bf B 136}, 
              115 (1978);
              B. Pendleton and G. G. Ross, Phys. Lett. {\bf B 98}, 291 (1981);
              C. Hill, Phys. Rev. {\bf D 24}, 691 (1981);
              G. M. Asatryan, A. N. Ivannissyan, and S. G. Matinyan, Yad. Fiz. 
              {\bf 53}, 592 (1991).
\bibitem{bc} H. Arason, D. Castano, B. Keszthelyi, S. Mikaelian, E. Piard,
             P. Ramond, and B. Wright, Phys. Rev. Lett. {\bf 67}, 2933 (1991).
\end{references}
\end{document}